\documentclass[superscriptaddress,12pt,showpacs]{revtex4-1}

\usepackage{amsmath}
\usepackage{amssymb}
\usepackage{graphicx}

\newcommand{\ee}{\mathrm{e}}

\newcommand{\TcJL} {{T_\mathrm{c}^J(L)}}
\newcommand{\TcJLR} {{T_\mathrm{c}^J(L)}}
\newcommand{\DTcL} {{\Delta T_\mathrm{c}(L)}}
\newcommand{\DTcN} {{\Delta T_\mathrm{c}(N)}}

\newcommand{\cSG} {\chi_{SG}}

\begin{document}

\title{Finite-Size Scaling Analysis of the Distributions of Pseudo-Critical
  Temperatures in Spin Glasses}
\author{A. Billoire}
\affiliation{Institut de physique th\'eorique, CEA Saclay and CNRS, 91191
  Gif-sur-Yvette, France.} 
\author{L.A. Fernandez} 
\affiliation{Departamento de F\'\i{}sica Te\'orica I,
  Universidad Complutense, 28040 Madrid, Spain.}  
\affiliation{Instituto de
  Biocomputaci\'on y F\'{\i}sica de Sistemas Complejos (BIFI), 50018 Zaragoza,
  Spain.}  
\author{A. Maiorano} 
\affiliation{Dipartimento di Fisica,
  Sapienza Universit\`a di Roma, P. A. Moro 2, 00185 Roma, Italy} 
\author{E. Marinari} 
\affiliation{Dipartimento di Fisica,
  Sapienza Universit\`a di Roma, P. A. Moro 2, 00185 Roma, Italy} 
\author{V. Martin-Mayor} 
\affiliation{Departamento de F\'\i{}sica
  Te\'orica I, Universidad Complutense, 28040 Madrid, Spain.}
\affiliation{Instituto de Biocomputaci\'on y F\'{\i}sica de Sistemas Complejos
  (BIFI), 50018 Zaragoza, Spain.}  
\author{D. Yllanes}
\affiliation{Departamento de F\'\i{}sica Te\'orica I, Universidad Complutense,
  28040 Madrid, Spain.}  
\affiliation{Instituto de Biocomputaci\'on y
  F\'{\i}sica de Sistemas Complejos (BIFI), 50018 Zaragoza, Spain.}  \date{
  \today}

\begin{abstract}
Using the results of large scale numerical simulations we study the
probability distribution of the pseudo critical temperature for the
three dimensional Edwards-Anderson Ising spin glass and for the fully
connected Sherrington-Kirkpatrick model. We find that the behaviour of
our data is nicely described by straightforward finite-size scaling
relations.
\end{abstract}

\date{\today}

\maketitle

\section{Introduction}
A proper phase transition takes place only in the
idealised limit of an infinite number of interacting degrees of
freedom. Although this limit is never realised in the laboratory (let
alone in numerical simulations), everyday experience suggests that
macroscopic samples are infinite for all practical purposes. Spin
glasses~\cite{mydosh:93,fisher:93} are an exception. The problem lies
in their sluggish dynamics at the critical temperature and below. The
system remains for very long times, or forever, out of equilibrium. In
fact, letting the system relax for about one hour, the spatial size of
the glassy magnetic domains is (at most) of the order of one hundred
lattice spacings~\cite{ORBACH}.

It has become clear lately that, in order to interpret experimental
data in spin glasses, the relevant equilibrium properties are those of
systems of size similar to that of the experimentally achievable
coherence length~\cite{janus:10,janus:10b}. Phase transitions on
finite systems are actually crossover phenomena describable through
the well known theory of finite size scaling (see
e.g.~\cite{amit:05}). However, a conspicuous feature of disordered
systems (and most notably, of spin glasses) is to undergo strong
sample-to-sample fluctuations in many thermodynamic properties. It is
thus natural to ask questions about the probability distribution,
induced by the disorder, of the various physical quantities. Typically
the size of these fluctuations decreases when enlarging the size of
the equilibrated system; if we wish to have hints about their possible
relevance in experimental systems,  it is important to know the rate
at which fluctuations decrease with system size. This is particularly
important if we are to study dynamical
heterogeneities~\cite{BIROLI-BOUCHAUD} in spin
glasses~\cite{janus:10b}. In particular, a relevant but elusive
physical quantity (potentially relevant to analyse dynamical effects
close to the phase transition) is the finite-system pseudo-critical
temperature. Our scope here is to characterise its statistical
properties in spin glasses.

This problem has been extensively studied and is well understood
for finite-size weakly bond-disordered spin models, below the upper
critical dimension $d_\text{up}$. For a system of size $N=L^d$ and a disorder sample $J$,
one can define a pseudo-critical temperature $\TcJL$ as the location
of the maximum of a relevant susceptibility: this definition is
clearly non unique, but all sensible definitions lead to the same
scaling behaviour as $N\to\infty$. According to the Harris
criterion~\cite{Harris}, a major role~\cite{AH,Paz1,Domany,AHW,Paz2}
is played here by the value of the thermal critical exponent of the
pure system, $\nu_P$. If $\nu_P>2/d$ the disorder is irrelevant, the
value of $\nu$ is not modified by the disorder (i.e. $\nu=\nu_P$), and
the width $\DTcL$ of the probability distribution of the
pseudo-critical temperature, defined as $\DTcL^2 \equiv
E(\TcJL^2)-E(\TcJL)^2$, where $E(\cdots)$ denotes the disorder
average, behaves as $\DTcL\propto 1/L^{d/2}$ as expected
naively~\cite{Domany95}. In such a situation in the infinite volume
limit the (disorder induced) fluctuations of $\TcJL$ are negligible
with respect to the width of the critical region and to the
finite-size shift of $T_\mathrm{c}$, that both behave like
$L^{-1/\nu_P}$. In the other case, when $\nu_P<2/d$. disorder is
relevant, the value of $\nu$ for the disordered model is different
from $\nu_P$ and obeys~\cite{Chayes} the bound $\nu>2/d$. In this case
$\DTcL$, the width of the critical region, and the finite-size shift
of $T_\mathrm{c}$ behave like $L^{-1/\nu}$. The behaviour $\DTcL\propto
1/L^{d/2}$ that would be naively dominant is destroyed by the
disorder. The case of weakly bond-disordered spin models above the
upper critical dimension needs a very careful analysis, as shown
in~\cite{Sarlat}.

To the best of our knowledge, the distribution of the pseudo-critical
temperature in finite size spin glass models has not been studied
numerically before: this is the object of the present note.  Recent
analytical work has predicted $\DTcN\propto 1/N^{2/3}$ (where $N$ is
the number of spins, i.e. the system volume) for the (mean field)
Sherrington-Kirkpatrick model (SK) for spin glasses~\cite{Castellana}:
we establish in this note that the realised scenario is indeed
different. Very recently, while this work was being completed,
ref.~\cite{Castellana2} has also tried (and failed) to verify
numerically the analytical predictions of~\cite{Castellana}.  
A former attempt to analyse numerically the
distribution of the pseudo-critical temperature in the SK model was
useful to investigate the numerical techniques of
choice~\cite{coluzzi:03}.

Here, we present numerical results both for the three-dimensional
($3d$) and the mean-field SK spin glass models.  In the $3d$ case, we
show that the probability distribution of pseudo-critical temperatures
verifies finite size scaling. From the scaling of this distribution we
obtain a precise estimate of the critical temperature and of the
critical exponent for the correlation length, $\nu$. On the other
hand, we find that for the mean-field spin glass $\DTcN\propto
1/N^{1/3}$ (in agreement with analytical findings for the scaling with
$N$ of disordered-averaged quantities in mean field
models~\cite{parisi:93}). Since this is in plain contradiction with
the results of Ref.~\cite{Castellana}, we briefly revisit their
analytical argument and show where the error in~\cite{Castellana}
stems from.  We also believe that a second analytic conclusion
of~\cite{Castellana}, stating that the $\DTcN$ is distributed
according to a Tracy Widom probability law, is based on very shaky
grounds, and we will give hints of the fact that it is not
substantiated numerically.

Our first step is to define a pseudo-critical temperature for a given
finite-size sample. Random bond and site diluted models
allow~\cite{Paz1,Domany,AHW,Paz2} a straightforward definition of
$\TcJL$ as the location of the maximum of the relevant susceptibility
$\chi$.  In our case the situation is more complex (even if, as we
will see, the analysis of the spin glass susceptibility $\cSG$ will be
very useful and revealing). Here the relevant diverging quantity
is~\cite{Foot1} $\cSG$
\begin{equation} 
  \chi^J_\mathrm{SG} =
  \frac{1}{N}\ \sum_{x,y} ( \langle S_x S_y\rangle_J -\langle
  S_x\rangle_J\langle S_y\rangle_J) ^2\;,
\end{equation}
(where the $S_x$ are the local spin variables)
that is of order $N$ in the whole low temperature phase. $\cSG$ is a
continuously decreasing function of the temperature, and has no peak
close to $T_\mathrm{c}$: this requires, as we will discuss in the
following, a slightly more sophisticated analysis in order to extract
a pseudo critical temperature.

An alternative and simpler procedure is very straightforward: let us
introduce it first.  We first assume (as done
in~\cite{AH,Paz1,Domany,AHW,Paz2}) that for a given disorder sample
$J$ the finite-size scaling of an observable $P$ of dimension $\zeta$
is $\langle P\rangle_J\simeq L^{\zeta}\;F((T-\TcJLR)L^{1/\nu})$, where
$F(\cdot)$ is a $L$ and $J$ independent finite-size scaling function:
the whole disorder sample dependence is encoded inside the
pseudo-critical temperature $\TcJLR$. This is in fact an approximation
since the scaling function has a residual $J$
dependence~\cite{Domany}.  We next build dimensionless combinations of
operators: we call them $O^J(T,L)$, and they are build in such a way
to scale as
\begin{equation} 
\label{FSS}
O^J(T,L)\simeq G((T-\TcJLR)L^{1/\nu})\;, 
\end{equation}
where $G(\cdot)$ is a $T$, $L$ and $J$ independent finite size scaling
function.  A familiar looking combination is the single sample
pseudo-Binder cumulant $B^J \equiv {\langle q^4\rangle_J} / {\langle
  q^2\rangle_J^2}$ (notice that this is defined for a given disorder
realisation): the genuine Binder cumulant is defined as $B \equiv
{E(\langle q^4\rangle)} / {E(\langle q^2\rangle)^2}$.  For sensible
choice of $y$ the solution $T_y^J$ of the equation $O^J(T_y^J,L)=y$,
with a disorder independent constant $y$, is a proxy of the
pseudo-critical temperature, namely $T_y^J = \TcJLR + C _y \;
L^{-1/\nu}=T_\mathrm{c}+D _y L^{-1/\nu}$ with an $L$ and $J$
independent constant $C_y$.  For example for a function $O$ that in
the infinite volume limit is zero in one phase and one in the other
phase any constant $y$ in the interval $[0,1]$ will do: it is wise, in
order to minimise the corrections to scaling to choose a legitimate
value for $y$ such that $T_y^J$ is typically inside the critical
region (the value of $T_y^J$ depends on $y$ and on the choice made of
a dimensionless combination $O^J$).

We are also able to use $\chi_{SG}$ for determining $\TcJL$: in this
way we are able to monitor a quantity that diverges in the infinite
volume limit, and to use it to extract a pseudo-critical
temperature. The approach used to define $\TcJL$ in this case is based
on the same technique: we compare the single sample spin glass
susceptibility to a value close to the average spin glass
susceptibility at the critical temperature on a given lattice
size. This measurement is a good proxy for the direct measurement of
the position of an emerging divergence.

We have applied these ideas to the Edwards-Anderson model in $3d$ and
to the SK model. We used existing data obtained by the Janus
collaboration on systems with $L=8$ to $L=32$ for the $3d$ EA
model~\cite{janus:10}, and from~\cite{ABMM} for the SK model with $N$
ranging from $64$ up to $4096$. In both cases the quenched random
couplings can take the two values $\pm 1$ with probability
one half.

The layout of the rest of this work is as follows. In
Sect.~\ref{sect:EA} we discuss our numerical methods, and we present
our results for the Edwards-Anderson model. An analogous analysis for
the SK model is presented in Sect.~\ref{sect:SK}. This study is
complemented in Sect.~\ref{sect:TAP} with our analysis of the
analytically predicted scaling for the distribution of pseudo-critical
temperatures. We also present in Sect.~\ref{sect:PD} an analysis of
the distribution function of the pseudo critical points.  Finally, we
give our conclusions in Sect.~\ref{sect:conclusions}.

\section{The Edwards-Anderson $3d$ model}\label{sect:EA}

The Hamiltonian of the model is
\begin{equation}
H_{3d}\equiv - \sum_{{\boldsymbol x},{\boldsymbol y}}
 S_{\boldsymbol x}\, J_{{\boldsymbol x},{\boldsymbol y}}\,
 S_{\boldsymbol y}\; ,
\end{equation}
where the sum runs over the couples of first neighbouring sites of a
$3d$ simple cubic lattice with periodic boundary conditions. The Ising $S_{\boldsymbol x}$ spin
variables can take the two values $\pm 1$ and the couplings are
quenched binary variables that can take the value $\pm 1$ with
probability one half.

In order to analyse the single sample pseudo-critical temperatures of
Edwards-Anderson $3d$ systems we need to construct several
dimensionless quantities.
We define the Fourier Transform of the replica-field $q_{\boldsymbol
  x}=S_{\boldsymbol x}^a S_{\boldsymbol x}^b$ ($S^a$ and $S^b$ are two
real replicas, i.e. two independent copies of the system evolving
under the same couplings, but with different thermal noise)
\begin{equation}
\phi(\boldsymbol k) = \sum_{\boldsymbol x} q_{\boldsymbol x}
\ee^{\mathrm i \boldsymbol k\cdot \boldsymbol x}\;,
\end{equation}
which we use to construct the two-point propagator
\begin{equation}
G^J(\boldsymbol k) = \langle \phi(\boldsymbol k) \phi(-\boldsymbol
k)\rangle_J\;,
\end{equation}
where $\langle\cdots\rangle_J$ denotes the thermal average for the
sample $J$.  Since the smallest momentum compatible with the periodic
boundary conditions is $|k| = 2\pi/L$ we define
\begin{align}
\boldsymbol k_{1}^{(1)} &= \bigl( 2\pi/L,\ 0,\ 0\bigr)\;, &
\boldsymbol k_{1}^{(2)} &= \bigl( 0,\ 2\pi/L,\ 0\bigr)\;, &
\boldsymbol k_{1}^{(3)} &= \bigl( 0,\ 0,\ 2\pi/L\bigr)\;,
\end{align}
and
\begin{equation}
G^J(\boldsymbol k_1) = \frac13 \sum_{i} G^J(\boldsymbol k_1^{(i)})\;.
\end{equation} 
Similarly, the second smallest momentum is given by
$(2\pi/L,2\pi/L,0)$ and by the two other possibilities:
we use it to define $G(\boldsymbol k_2)$.

We consider the following dimensionless quantities:
\begin{align}
\xi^J/L & \equiv \frac{1}{2L \sin(\pi/L)} \left[
  \frac{G^J(0)}{G^J(\boldsymbol k_1)} - 1\right]^{1/2}\;,\label{eq:xi}
\\ B^J &\equiv \frac{\langle q^4\rangle_J}{\langle q^2\rangle^2_J}\;,\label{BJ}
\\ B_G^J
&\equiv \frac{ \sum_i \bigl\langle\bigr[\phi(\boldsymbol k_1^{(i)})
    \phi(-\boldsymbol
    k_1^{(i)})\bigr]^2\bigr\rangle_J}{[G^J(\boldsymbol
    k_1)]^2}\;,\\ 
R_{12}^J &\equiv \frac{G^J(\boldsymbol
  k_1)}{G^J(\boldsymbol k_2)}\;.\label{eq:R12}
\end{align}
In Eq.~\ref{BJ}
 we have used the global spin-overlap, which is defined as
\begin{equation}
q=\frac{\phi({\boldsymbol k}=0)}{N}\,,
\end{equation}
where $N=L^d$ is the total number of spins.
Let us start by considering the sample-averaged observables (which we
denote by dropping the super-index $J$). Up to scaling corrections,
they do not depend $L$ at the critical point,
\begin{equation}
O(T_\text{c}, L) = y_\text{c} + O(L^{-\alpha})\;,\mbox{ with } \alpha>0\;.
\end{equation}
\begin{figure}
\centering
\includegraphics[height=0.7\linewidth,angle=270]{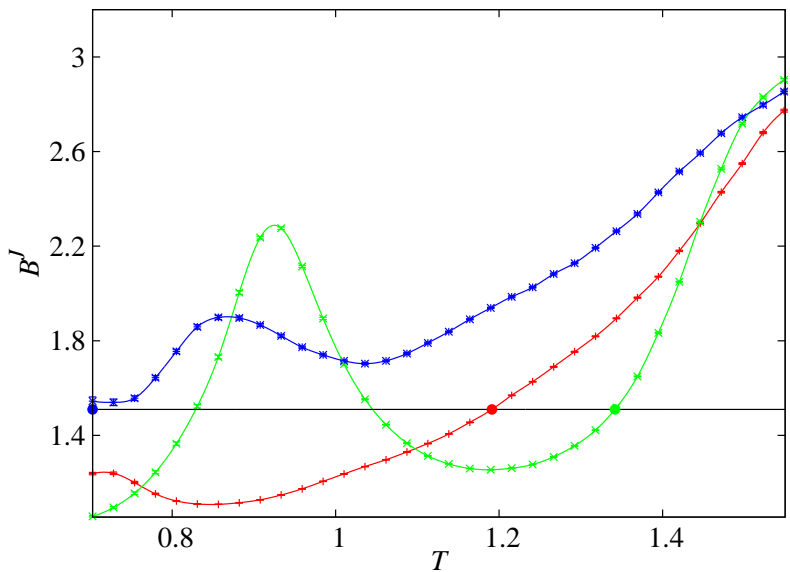}
\caption{Computation of the pseudo-critical temperature using the
  Binder ratio defined in eq.~\ref{BJ}.  Here $y=1.51$, that is a good
  approximation to $B(T_\text{c})$, and we plot $B^J(T)$ for three
  samples with $L=32$.  The red curve has exactly one solution (in our
  range) for $B^J(T) = y$, which defines its $T_{y=1.51}^J$: this is
  the normal behaviour of this observable.  The green sample has three
  solutions, so we pick the largest one (green dot).  The blue sample
  has $B^J(T)>y$: in this case we take the lowest simulated
  temperature $T\approx0.7$ as an upper bound for the pseudo-critical
  temperature (this ignorance will not affect our estimate for the
  median). Error bars are included in the plot and are very small.
\label{fig:TcJ}}
\end{figure}
For each value of $L$ we can search for the temperature
$T_{c,y}^L$  such that 
\begin{equation}
O(T_{c,y}^L ,L) = y\;.
\end{equation}
Then, provided we are not very far from the scaling region (so that
$y$ is not too different from $y_\text{c}$), we expect that
\begin{equation}\label{eq:Ty}
T_{c,y}^L  \simeq T_\text{c} + A_y L^{-1/\nu} ( 1+ B_y L^{-\omega})\;,
\end{equation}
where we have included the first corrections to scaling.

We can use this same approach to define a single-sample critical
temperature $T_\text{c}^J$.  Let us choose a fixed value of $y$ close
enough to the value $y_\text{c}$ defined from the sample average at
the critical point. For each sample we use cubic splines to
determine $T_y^{L,J}$ such that
\begin{equation}
\label{eq:TcJ}
O^J( T_y^{L,J},L) = y\;.
\end{equation}
For some samples the $O^J(T,L)$ turn out not to be monotonic: there can be
several solutions to this equation. In those cases we simply pick the largest
solution. This process is illustrated in Figure~\ref{fig:TcJ}.

The motivation for this choice is simple.  The physical meaning of a
pseudo-critical temperature is a characteristic temperature that separates the
paramagnetic phase from the low-temperature one. Indeed, any temperature
$T_y^J$, solving the equation $O^J(T_y,L)=y$, is a temperature where
non-paramagnetic behaviour has already arisen. Therefore, only the largest
$T_y^J$ makes sense as a divider among both phases. In fact, the non-monotonic
behaviour of $O^J(T,L)$ may be due to other reasons (in particular,
temperature chaos), unrelated to the paramagnetic/spin-glass phase transition.
In any case, our definition will be justified \emph{a posteriori}, on the
view of the simplicity of the emerging physical picture.

The values of $T_y^J$ have a very wide probability distribution.  For
a few disorder samples the solution of eq.~\eqref{eq:TcJ} falls
out of our simulated range of temperatures and we only obtain an upper
or (less frequently) a lower bound (see the blue curve in
Figure~\ref{fig:TcJ}).  In this situation the arithmetic average of the
$T_y^J$ is not well defined: we consider instead the median
temperature, that we denote by $\tilde{T}_{y}^J$.
Since, by definition, the median does not change  as long as the proportion 
of samples without a solution is less than 50\%, this
is a robust estimator in these circumstances 
(we are well below this limit for all
the cases considered, the typical proportion being about $\sim1\%$).
  From fig.~\ref{fig:TcJ} it is
also clear that the statistical uncertainty over the determination of
$T_y^J$ in a given sample is very small as compared to the size of
sample to sample fluctuations.

In analogy with the sample averaged case~\eqref{eq:Ty}, we make the
ansatz
\begin{equation}
\label{eq:fitM}
\tilde{T}_{y}^J(L) \simeq T_\text{c} + A_y L^{-1/\nu}\;,
\end{equation}
where we have ignored sub-leading corrections.  A fit to this equation
would, in principle, yield the values of $T_\text{c}$ and
$\nu$. However, for a fixed value $y$, we do not have enough degrees
of freedom to determine simultaneously $\nu$ and $T_\text{c}$.

Following the approach of~\cite{janus:10b}, we get around this problem
by considering $n$ values of $y$ at the same time: this allows us to fit
at the same time all the resulting $\tilde{T}_y^J(L)$, with fit
parameters $\{T_\text{c}, \nu, A_{y_1},\ldots,A_{y_n}\}$: in other
words we force the $\tilde T_y^J$ obtained for different $y$ values to
extrapolate to the same $T_\text{c}$ with the same exponent. This
procedure may seem dangerous, since we are extracting several
transition temperatures from each of the $O^J(T,L)$, that are
correlated variables. However, the effect of these correlations can
be controlled by considering the complete covariance matrix of the
data.

The set of points $\{\tilde T_{y_i}^J(L_a)\}$ are labelled by their $L$
and their $y$: we have data for ${\cal L} = 5$ different values of
$L$, with $L_1=8$, $L_2=12$, $L_3=16$, $L_4=24$, $L_5=32$ ($4000$
samples in all cases but for $L=32$, where we have $1000$).  We also
select $n$ values of $y_i$ in the critical region ($y_1$, $\ldots$,
$y_n$). The appropriate chi-square estimator is
\begin{equation}
\chi^2 = \sum_{i,j=1}^n \sum_{a,b=1}^{\cal L}
\bigl[ \tilde T_{y_i}^J(L_a) - T_\text{c} - A_{y_i} L_a^{-1/\nu}\bigr]
\sigma^{-1}_{(ia)(jb)}
\bigl[ \tilde T_{y_j}^J(L_b) - T_\text{c} - A_{y_j} L_b^{-1/\nu}\bigr]\;,
\end{equation}
where $\sigma_{(ia)(jb)}$ is the covariance matrix of the set of
$T_y^J$, which we compute by a bootstrap approach~\cite{boostrap}
(each point of the set is identified by $L$ and $y$): it is a
block-diagonal matrix, since the data for different $L$ values are
uncorrelated.

Thus far we have considered just a single observable $O$, for
different values of $y$. Since the fitting function~\eqref{eq:fitM} is
the same for the different $O$ we have selected, with common
$T_\text{c}$ and $\nu$, and only amplitudes differ, we can include in
the same fit data for the four dimensionless quantities
(\ref{eq:xi}--\ref{eq:R12}), considering several values of $y$ for
each. In order to simplify the notation, from here on we shall denote
our set of points in the fit as $\{\tilde T_\alpha^J(L)\}$, taking
$\alpha$ as labelling both the observable $O$ and the height $y$ (so
that it will range from $1$ to $4\,n$).

We can use the usual disorder averaged spin-glass susceptibility
$\chi_\mathrm{SG} $ to arrive at yet another determination of the
single-sample critical temperature, with the definition
\begin{equation}
\chi_\text{SG}^J(T^J_\chi) = \chi_\mathrm{SG} (T_\mathrm{c})\; y\;,
\end{equation}
with $y$ close to one, and we expect $\tilde T^J_\chi$ to follow the
same scaling behaviour of~\eqref{eq:fitM}: we include the values of
$\tilde T^J_\chi$ in the global fit to the individual pseudo critical
temperatures. In this case the pathologies that affect the analysis
of other single sample observables are far less frequent.

\begin{figure}
\includegraphics[height=0.7\linewidth,angle=270]{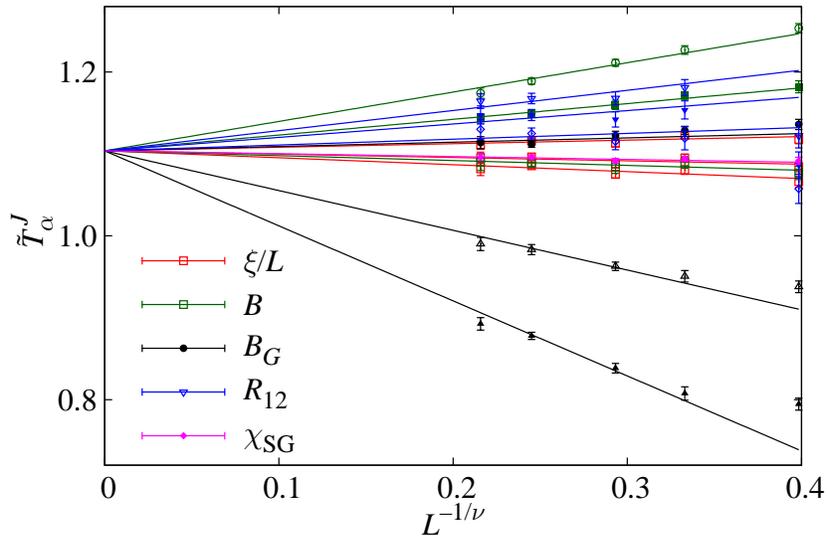}
\caption{Global fit for the median single-sample critical temperature
  $\tilde T_\alpha^J$ computed with different observables.  The
  parameters in the fit are the common extrapolation point
  $T_\text{c}$, the common exponent $\nu$ and an individual amplitude
  for each set of data points. We include in the fit the data for
  $L\geq 12$.
\label{fig:fitM}}
\end{figure}

We show the results of this combined fitting procedure in
Figure~\ref{fig:fitM}, where we have included data for the four
dimensionless ratios $\xi/L$, $B$, $B_G$ and $R_{12}$, using three
values of $y$ for each ratio (we plot the three fits for the same
observable with the same colour). We have also used the data for
$T_\chi^J$, with $y=1$ (we select the value of $T_\text{c}$ reported
in~\cite{hasenbusch:08b}).  We have discarded the $L=8$ data, which
showed strong corrections to the leading scaling
of~\eqref{eq:fitM}. The best fit gives
\begin{align}
T_\text{c} &= 1.104(6)\;, & \nu&=2.26(13)\;,
\end{align}
with $\chi^2=32.6$ for $37$ degrees of freedom (giving a $P$ value of
$68\%$).  This results nicely agree with the determination
of~\cite{hasenbusch:08b},
\begin{align}
T_\text{c} &= 1.109(10), & \nu &= 2.45(15).
\end{align}
Ref.~\cite{hasenbusch:08b} includes corrections to scaling, as 
in~\eqref{eq:Ty}, with $\omega=1.0(1)$.
Ref.~\cite{K1} gives a comprehensive list of estimates for $T_c$ and $\nu$.
In order to take these scaling corrections into account and to include the
$L=8$ data in the fit, we can therefore rewrite~\eqref{eq:fitM} as
\begin{equation}\label{eq:fitM-corr}
\tilde T_{\alpha}^J(L) \simeq T_\text{c} + A_\alpha L^{-1/\nu}(1 +
B_\alpha L^{-\omega}),
\end{equation} 
where again we use the same $\omega$ parameter for all the observables
and all values of $y$. Unfortunately, our numerical data are not
precise enough to allow a reliable determination of $\omega$, $\nu$
and $T_\text{c}$ at the same time (the resulting error in $\omega$
would be greater than $100\%$).  We have been able to check
consistency of our approach by taking the values of $\nu$ and $\omega$
from~\cite{hasenbusch:08b} and fitting only for $T_\text{c}$ and for
the amplitudes, including now the data for $L=8$.  The resulting best
fit gives $T_\text{c}=1.105(8)$, with $\chi^2 = 41.9$ for $38$ degrees
of freedom ($P$ value: $35\%$): this is a satisfactory check of
consistency. 

The results we have discussed make us confident of the fact that our
determination of the single-sample critical temperatures yields
reasonable results. We can now take the analysis one step further and
consider the width of the distribution of $T_\text{c}^J$.  We consider
the two temperatures $T^+_\alpha$ and $T^-_\alpha$ such that
\begin{align}
P(T_\alpha^J > T_\alpha^+ ) &= 0.16\;,&
P(T_\alpha^J < T_\alpha^- ) &= 0.16\;.
\end{align}
The value $0.16$ is such that the temperature interval
$[T_\alpha^-, T_\alpha^+]$ defines the same probability as an
interval of two standard deviations around the mean for a Gaussian
probability distribution.  We define the width $\Delta T_\alpha^J$ as
\begin{equation}
\Delta T_\alpha^J = \frac{T^+_\alpha-T^-_\alpha}2\; .
\label{width}
\end{equation}
\begin{figure}
\includegraphics[height=0.7\linewidth,angle=270]{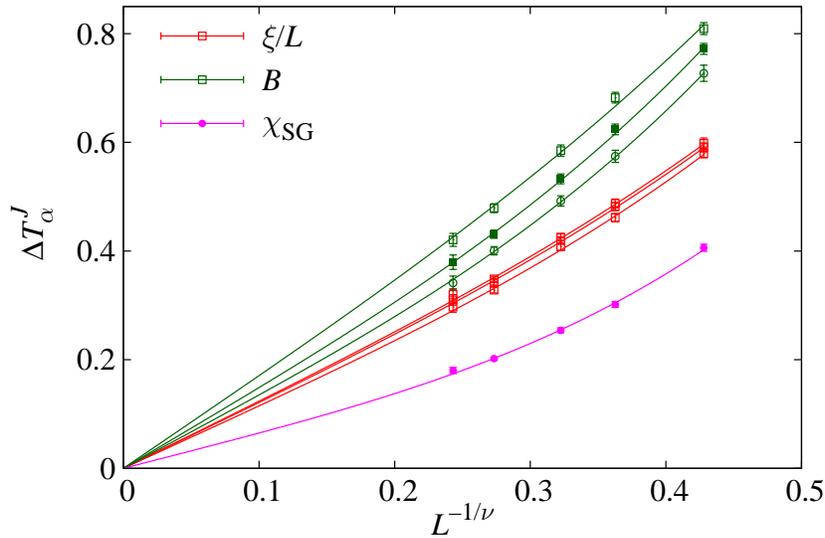}
\caption{Global fit to $\Delta T_\alpha^J$, computed
  using $\xi^J/L$, $B^J$ and $\chi_\text{SG}^J$.
\label{fig:fitwidth}}
\end{figure}
The simplest ansatz for the scaling behaviour of $\Delta T_\alpha^J$ is
\begin{equation}\label{eq:fitDelta}
\Delta T_\alpha^J \simeq A_\alpha L^{-1/\nu} ( 1+ B_\alpha L^{-\omega})\;.
\end{equation} 
In principle we could repeat the global fitting procedure that we have
applied to the medians $\tilde T_\alpha^J$. Unfortunately, not all the
observables that we considered in Figure~\ref{fig:fitM} can be used to
analyse $\Delta T_\alpha^J$, since the distribution of some of them is
too wide, so that the critical temperature of too many samples falls
out of our simulated range of $T$, and the width defined in
eq.~\ref{eq:fitDelta} is undefined.  Because of that we analyse
$\Delta T_\alpha^J$ by only using the $T_y^J$ derived from $\xi^J/L$,
$B^J$ and $\chi_\text{SG}^J$.  The corrections to scaling are now
stronger than for the median $\tilde T_y^J$, so that we cannot obtain
a good fit to leading order even if we discard the data for $L=8$.
Using once again as an input the critical exponents
from~\cite{hasenbusch:08b} and fitting for the amplitudes we obtain a
very good fit with $\chi^2=19.1$ for $21$ degrees of freedom ($P$
value: $58\%$).  The results of the best fit are plotted in
Figure~\ref{fig:fitwidth}.  According to the ansatz of
eq.~\ref{FSS}, the width of $T_y^J$ must be equal to the width of
$\TcJLR$ (since $C _y$ is $J$ independent): the width of $T_y^J$
should accordingly be $y$ independent.  Figure~\ref{fig:fitwidth}
shows that this is indeed the case for $\Delta T_{\xi/L}^J$, but less
so for $\Delta T_B^J$: as mentioned before the disorder independence
of the scaling function in eq.~\ref{FSS} is only
approximate~\cite{Domany}.

\section{The Sherrington-Kirkpatrick mean field theory}\label{sect:SK}

The Hamiltonian of the Sherrington-Kirkpatrick mean field model is
\begin{equation}
H_\text{SK}\equiv -\frac{1}{\sqrt{N}} \sum_{i,j}
 S_{i}\, J_{i,j}\,
 S_{j}\; ,
\end{equation}
where the sum runs over all couples of spins of the system, the Ising
$S_{i}$ spin variables can take the two values $\pm 1$ and
the couplings are quenched binary variables that can take the value
$\pm 1$ with probability one half.

\subsection{The pseudo-critical temperatures}\label{sect:SKDATA}

Our analysis of the mean field SK model is very similar to the one we
have discussed in the case of the Edwards-Anderson model. Here we have
seven values of the system size $N$ (of the form $N=2^p$, with $p$
ranging from $6$ to $12$: in all cases we have $1024$ disorder samples
but for $N=4096$ where we have $256$ and for $N=128$ where we have
$8192$), that makes the fitting procedure stable.  It is also of use
the fact that in this case the value of the (infinite volume) critical
temperature, $T_c=1$ is known exactly.

We consider three definitions of $T_\mathrm{c}$:
\begin{itemize}
\item the one based on the single sample Binder cumulant of
  eq.~\ref{BJ}, $B^J(T^J_B)=B(T_\mathrm{c})$;

\item one based on the low order cumulant $D=E(\langle
  q^2\rangle)/E(\langle|q|\rangle)^2$ and the single sample quantity
  $D^J = {\langle q^2 \rangle_J} / {\langle |q|\rangle^2_J}$, i.e.
  $D^J(T^J_D)=D(T_\mathrm{c})$;

\item one based on the spin-glass susceptibility
  $\chi_{SG}^J(T^J_\chi) = \chi_\mathrm{SG} (T_\mathrm{c})$.
\end{itemize}
For a given value of $N$, we use for the left hand sides
$B(T_\mathrm{c})$, $D(T_\mathrm{c})$ and $\chi_\mathrm{SG}
(T_\mathrm{c})$ the values measured for this same size, at
$T_\mathrm{c}=1$. We solve eq.~\eqref{eq:TcJ} by using a simple linear
interpolation.

Like for the EA model, in some cases eq.~\eqref{eq:TcJ} can have more
than one solution. We again choose the largest solution, which in the
case of SK turns out to be always the one closer to the infinite
volume value $T_\mathrm{c}$.  In a few cases, for small values of $N$,
the equation has only solutions outside the range of temperatures that
was used in the parallel tempering Monte Carlo simulation ($0.4 \leq T
\leq 1.30$).  We fix this problem again by basing our statistical
analysis on the median of the distribution and on the definition of
the width given by eq.~\ref{width}. It turns out that these
pathological cases are less numerous for the SK model than for the EA
model, and that the width given by eq.~\ref{width} is always defined.

\begin{figure}
\includegraphics[height=0.7\linewidth,angle=270]{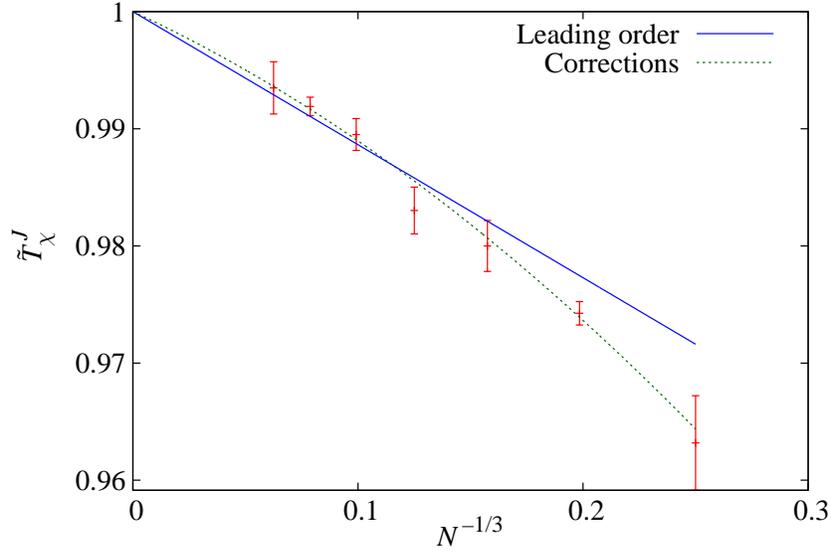}
\caption{The median of the distribution of $T^J_{\chi}$ as function of
  $1/N^{1/3}$, together with the results of two best fits: a leading
  order fit, and a fit including a higher order correction (see
  text).}
\label{fig:fitMSK}
\end{figure}

In terms of the number of sites $N$ of the SK fully connected lattice
the ansatz of eq.~\ref{eq:fitM} becomes
\begin{equation}
\tilde T_{y}^J(L) \simeq T_\text{c} 
+ A_y N^{-1/(\nu d_\text{up})}
=T_\text{c} + A_y N^{-1/3}\;.
\end{equation}
We show in fig.~\ref{fig:fitMSK} the data for the median of the
distribution of $T^J_{\chi}$ as a function of $1/N^{1/3}$, together
with the results of two best fits.  We first notice that the data are
well compatible with the fact that in the $N\to\infty$ limit $T_c=1$.
The first fit is a linear fit to the form $T^J_{\chi}=1+a/N^{1/3}$
(with $N\ge 256$ and $\chi^2=4.44$ with $4$ degrees of freedom).  This
is a good fit for the large systems, but it fails below $N=256$.  We
also show a (very good) best fit including the next to leading
corrections, with an exponent $2/3$ (including all $N$ values,
$\chi^2=2.07$ with $5$ degrees of freedom).  The analysis of the data
for $T^J_B$ and $T^J_D$ leads to the same conclusions: here however
the leading term ($\propto 1/N^{1/3}$) has a small coefficient and the
effect of the next to leading term is stronger.  In conclusion our
data are in excellent agreement with an asymptotic $1+{\cal O}
(1/N^{1/3})$ behaviour for the median of the distribution.

\begin{figure}
\includegraphics[height=0.7\linewidth,angle=270]{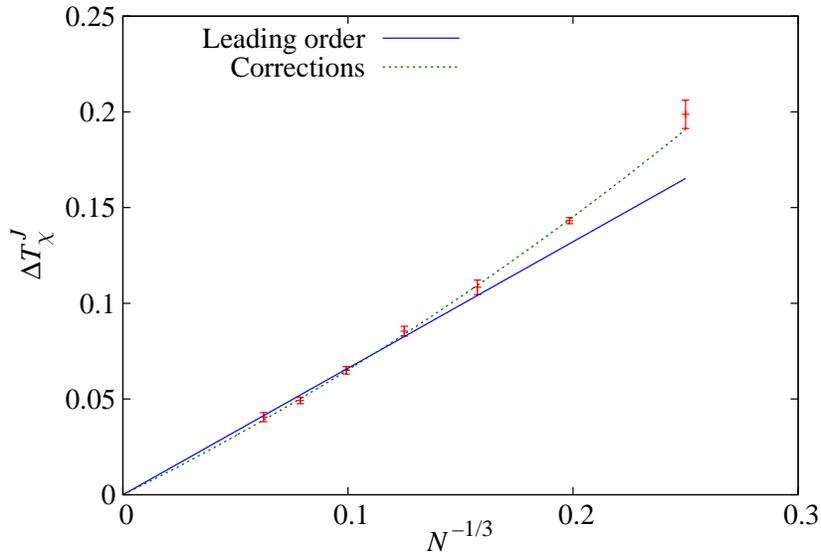}
\caption{ The width of the distribution of $T^J_{\chi}$ as
  function of $N^{-1/3}$, together with the results of two fits: a
  leading order fit, and a fit including a higher order correction
  (see text). }
\label{fig:fitwidthSK}
\end{figure}

We show in fig.~\ref{fig:fitwidthSK} the width of the distribution of
$T^J_{\chi}$ as a function of $1/N^{1/3}$, together with the results
of two fits, namely $\Delta T_\chi^J = c_1/N^{1/3}$, and $\Delta
T_\chi^J=c_1 N^{-1/3}+c_2N^{-2/3} $ respectively.  The data are well
compatible with $\Delta T=0$ in the limit $N\to\infty$, as
expected. The leading order fit, including $N\ge 256$, has a
$\chi^2=5.36$ with $4$ degrees of freedom.  The two-parameter fit
gives an excellent representation of the data (with a $\chi^2=2.475$
with $5$ degrees of freedom) including now the $N=64$ and $N=128$
points. Very similar results are obtained for $\Delta T_B^J$ and
$\Delta T_C^J$.

In conclusion our finite size, numerical analysis of the SK model
strongly support an asymptotic ${\cal O} (N^{-1/3})$ scaling behaviour
for the width of the distribution of the pseudo-critical temperatures.

\subsection{Scaling with the system size and the stability of TAP
  states}\label{sect:TAP}

Our results are in contradiction with the claim made
in~\cite{Castellana} that the width of the finite-size fluctuations
due to quenched disorder of the critical temperature of the SK spin
glass scales like $N^{-2/3}$. We show here that we can give support to
our numerical finding by means of a very simple scaling argument.

The SG susceptibility can be computed from the TAP 
free energy~\cite{Thouless:77} as
\begin{equation}\label{CHI-TAP}
\chi_\mathrm{SG}=\frac{1}{N} \mathrm{Tr} \mathcal{M}^{-2}\;,
\end{equation}
where ${\cal M}$ is the Hessian of the TAP free energy at the relevant
minimum.  If we seat deep in the paramagnetic phase, the only relevant
minimum of the TAP free energy is $m_i=0$ for all $i=1,2,\ldots,N$.
Note that in the pseudo-critical region, where $(\beta
-\beta_\mathrm{c})N^{1/3}\sim 1$, is not at all obvious that the such
TAP solution is the relevant one (for instance a sub-extensive set of
sites of size $N^\alpha$, with $\alpha<1$, could have non-vanishing
$m_i$, or maybe one could have for all sites $|m_i|\sim N^{-\alpha'}$,
with $\alpha'>0$): the following discussion is relevant only in
the paramagnetic phase and for system sizes so large that  
$(\beta_\mathrm{c}-\beta)\gg N^{-1/3}$.

It was shown some time ago~\cite{Mike} that, at $\beta =
\beta_\mathrm{c}$, the smallest eigenvalue of the Hessian at the
fully-paramagnetic TAP solution is of order $N^{-2/3}$. In
Ref.~\cite{Castellana} it has been argued that
\begin{equation}
\lambda_{\mathrm{min},J}=(1-\beta)^2+\frac{\beta}{N^{2/3}}
  \Phi^{(N)}_J\,,
\end{equation}
where $\Phi_J^{(N)}$ is a random variable that, in the limit
of large $N$, converges in distribution to a Tracy-Widom
random-variable~\cite{TW}. In particular, note that at the critical
temperature $\beta_\mathrm{c}=1$ the $N^{-2/3}$ scaling is recovered.

Now, only for the purpose of discussing the crudest features of the
scaling laws, let us assume that $\chi_\mathrm{SG}$ is dominated by
the contribution of the smallest eigenvalue:
\begin{eqnarray}
\chi_\mathrm{SG}^J&\sim& \frac{1}{N \lambda_{\mathrm{min},J}^2}\,,\\ &=&
\frac{1}{N\left( (1-\beta)^2+\frac{\beta}{N^{2/3}} \Phi^{(N)}_J \right)^2}\,,
\label{Eq:SK-very-bad}\\
&=& N^{1/3} \frac{1}{\left( \left[N^{1/3}(1-\beta) \right]^2 + \beta
  \Phi^{(N)}_J \right)^2}\,.\label{Eq:SK-bad}
\end{eqnarray}
The analysis of Ref.~\cite{Castellana} is based entirely on
eq.~(\ref{Eq:SK-very-bad}).

Now, note that eq.~(\ref{Eq:SK-bad}) implies that interesting behaviour appears
only when $(1-\beta)\leq 1/N^{1/3}$: this makes sensible to replace $ \beta
\Phi^{(N)}_J$ with $\Phi^{(N)}_J$. At this point, an implication emerges for
the scaling with $N$ of the average susceptibility. We have that
\begin{equation}
\chi_\mathrm{SG}(T,N)= N^{1/3} G\bigl( N^{1/3}(1-\beta)\bigr)\;,
\end{equation}
where the scaling function $G$ has the form
\begin{equation}\label{eq:G-def}
G(x)=\int_{-\infty}^{+\infty}\,\mathrm{d}\Phi\, p_\mathrm{TW}(\Phi)\,
\frac{1}{(x^2 +\Phi)^2}\,.
\end{equation}
In the above expression $p_\mathrm{TW}(\Phi)$ is the Tracy-Widom
probability density function. 

Eq.~(\ref{eq:G-def}) is not acceptable for two main reasons:
\begin{itemize}
\item the function $G(x)$ in eq.(\ref{eq:G-def}) is ill-defined, as
  the integrand has a non-integrable singularity at
  \hbox{$\Phi=-x^2$};
\item if one devises some regularisation procedure, dimensional
  analysis would indicate that $G(x)\sim x^{-4}$ in the limit of large
  $x$. However, in order to recover the correct critical divergence
  $\chi_\mathrm{SG}^{N=\infty}\sim 1/(1-\beta)$, one obviously needs
  $G(x)\sim 1/x$.
\end{itemize}
The solution of these two caveats is of course in the fact that the
initial assumption, $\chi_\mathrm{SG}^J \sim 1/(N
\lambda_{\mathrm{min},J}^2)$, is incorrect. The contribution of $\sim
N$ eigenvalues is crucial in order to recover the correct scaling
behaviour $G(x)\sim 1/x$. Thus, the only lesson that we may take from
this oversimplified analysis is that the $J$-dependent SG
susceptibility will probably scale as
\begin{equation}
\chi_\mathrm{SG}^J= N^{1/3}\mathcal{F}\left(\left[N^{1/3}
(1-\beta)\right]+\Psi_J^{(N)}\right)\,.
\end{equation}
This is exactly the scaling ansatz we made at the beginning, where
$\Psi_J^{(N)}$ is some random-variable that (in distribution) remains
of order $1$ in the large-$N$ limit. This result is consistent with
our numerical findings.

Merely rewriting eq.~(\ref{Eq:SK-very-bad}) as eq.~(\ref{Eq:SK-bad})
suffices to make it obvious that the asymptotic statement in
Ref.~\cite{Castellana} is incorrect: the width of the distribution of
the pseudo-critical temperatures scales with $N^{-1/3}$.  Indeed, if
as done in Ref.~\cite{Castellana}, one simply derives in
eq.~(\ref{Eq:SK-very-bad}) with respect to $\beta$ in order to get the
maximum of the susceptibility, one finds that
$1-\beta_{\mathrm{c},J}\sim N^{-2/3}$. But at such value of
$\beta_{\mathrm{c},J}$ we have that $(1-\beta)^2\sim N^{-4/3}\ll
\frac{\beta}{N^{2/3}} \Phi^{(N)}_J$, In other words, at the scale of
$(1-\beta)\sim 1/N^{2/3}$ eq.~(\ref{Eq:SK-very-bad}) predicts an
essentially constant behaviour, hence the supposed maximum of the
susceptibility (recall that at such value of $\beta$ the fully
paramagnetic TAP minimum is probably no longer the relevant one) has
no physical meaning.

\subsection{The probability distribution of the pseudo-critical inverse 
  temperatures}\label{sect:PD}

\begin{figure}
\includegraphics[height=0.7\linewidth,angle=0]{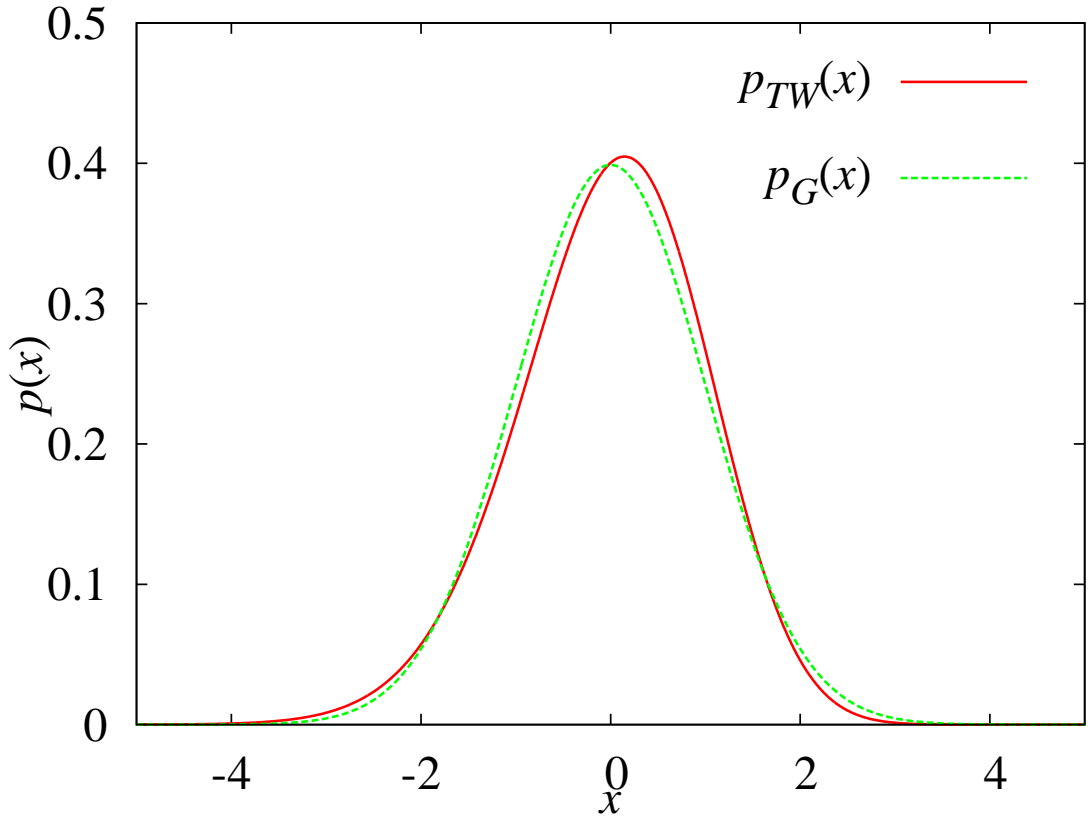}
\caption{Probability density functions for the Gaussian distribution and
  for the Tracy-Widom distribution for the Gaussian Orthogonal Ensemble,
  both with zero average and unitary variance.}
\label{fig:TW1}
\end{figure}

We have discussed in some detail about the features of the
pseudo-critical, sample dependent temperatures in the mean field SK
theory, and we have determined their scaling properties. The next
step, that we present here, is in the study of their probability
distribution. Unfortunately, there is not any clean analytical
prediction for the shape that this distribution function should take
in the large-$N$ limit. The only proposal known to us was put forward
in ref. \cite{Castellana}: when properly scaled, the pseudo-critical
temperatures should follow a Tracy-Widom (TW) distribution, in our
case for the Gaussian Orthogonal Ensemble (GOE). Unfortunately, as we
have explained in Sect.~\ref{sect:TAP}, the reasoning leading to that
prediction is flawed (although, on the long run, the prediction itself
could be correct).

Lacking an analytical guidance, we will simply check whether our
numerical data can be described by either a TW distribution, or by the
ubiquitous Gaussian distribution. Within our limited statistics and
system sizes, the two distributions turn out to be acceptable (the
Gaussian hypothesis fits slightly better our data, but a Tracy-Widom
hypothesis is certainly consistent as well). Given the preliminary
nature of this study, we shall restrict ourselves to the simplest
determination of pseudo-critical temperatures, the one coming from the
spin-glass susceptibility.

Let us start by noticing that the difference of a Gaussian
distribution $F_G(\phi)$ and a TW distribution for the GOE
$F_{TW}(\phi)$ is indeed very small. We show in figure \ref{fig:TW1}
both a Gaussian and a TW distribution with zero average and
variance equal to one: it is clear that they are very similar. Some
numerical values can be of help. In the case of zero average and
unitary variance a Gaussian has a fourth moment equal to $3$, as
opposed to $3.165$ for a TW. The Gaussian is symmetric and has
zero skewness, while the TW distribution has a small asymmetry, with a
skewness equal to -0.29, The Gaussian has a kurtosis equal to zero,
while a TW has a kurtosis equal to $0.165$. We will use these
numerical remark at the end of this section for sharpening the outcome
of our quantitative analysis.

It is clear that in this situation, where the two target distributions
are very similar, one has to keep under very strong control finite
size effects, that could completely mask the asymptotic behaviour.  It
is important to notice that the effects we are looking at characterise
not only the tails but the bulk of the distribution.

\begin{figure}
\includegraphics[height=1.1\linewidth,angle=0]{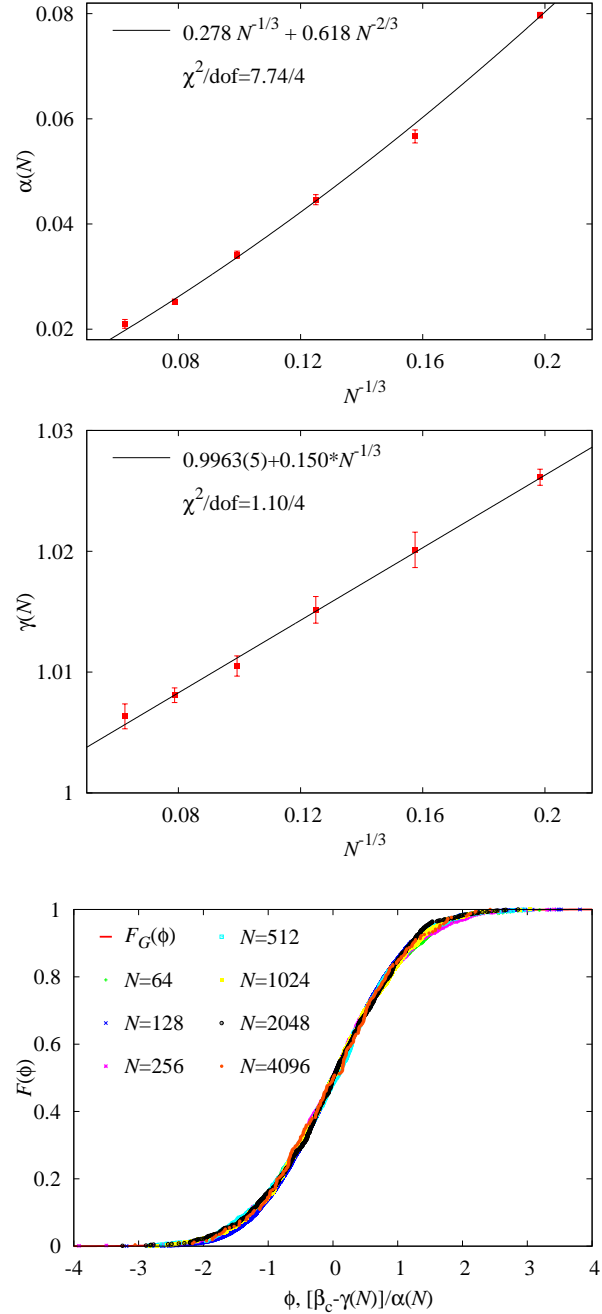}
\caption{ (top): $\alpha(N)$, defined in equation
  \ref{eq:wedoitbetter}, versus $N^{-1/3}$, and our best fit including
  the first scaling corrections.  (middle): $\gamma(N)$, defined in
  equation \ref{eq:wedoitbetter}, versus $N^{-1/3}$, and our best
  fit. Here the fitting function only includes the leading term, since
  this form already gives a good value for $\chi^2$.  (bottom):
  collapse of the EDF for different $N$ values and of the theoretical
  distribution function, as described in the text.  Here we show the
  case of a Gaussian distribution.}
\label{fig:TW2}
\end{figure}

\begin{figure}
\includegraphics[height=1.1\linewidth,angle=0]{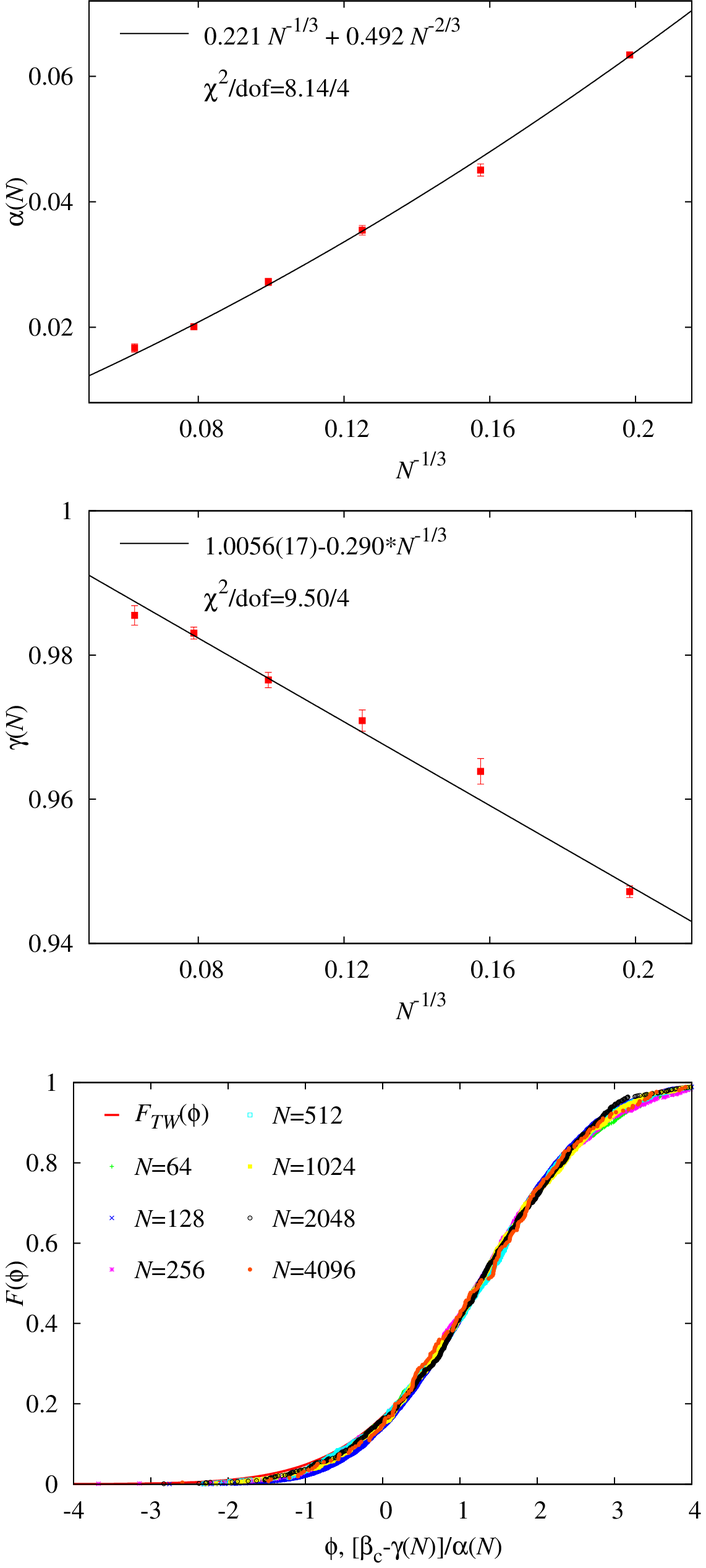}
\caption{As in figure \ref{fig:TW2}, but for a TW probability
  distribution.}
\label{fig:TW3}
\end{figure}

Let us give the basic elements of our approach. We consider the value
of the pseudo-critical inverse temperatures computed from the spin
glass susceptibility for different $N$ values. We try to verify if a
relation of the form
\begin{equation}
\label{eq:wedoitbetter}
  \beta_c^{(N)} = \gamma_N+\alpha_N\,\phi\;,
\end{equation}
(where $\phi$ is a random variable with null expectation values and
unit variance, Gaussian or TW like) can account for our numerical data
and, if yes, to determine the scaling behaviour of $\gamma_N$ and of
$\alpha_N$.  We now define the variable
$\tilde{\beta_\mathrm{c}}\equiv(\beta_c-\gamma)/\alpha$.
Let us assume that for a system of size $N$ we have $K_N$ samples,
and therefore $K_N$ pseudo-critical temperatures
$\tilde{\beta_\mathrm{c}}(N,s), s=1,2,\ldots,K_N$. The empirical
distribution function (EDF) is then
\begin{equation}
H_N(\tilde{\beta_\mathrm{c}}) \equiv \frac{1}{K_N} \sum_{s=1}^{K_N}
\theta(\tilde{\beta_\mathrm{c}}-\tilde{\beta_\mathrm{c}}(N,s))\,,
\end{equation}
where $\theta$ is the Heaviside step function.  Note that the
parameters $\alpha(N)$ and $\gamma(N)$ in Eq.~\ref{eq:wedoitbetter}
are unknown a priori. They will be determined through a fitting
procedure (see Eq.~\ref{eq:Dist1} below). In particular, they
typically take different numerical values if we adopt the TW
hypothesis, or the Gaussian hypothesis.

We define the distance
among the EDF and the theoretical distribution as
\begin{equation}
\label{eq:Dist1}
D\equiv \frac{1}{K_N} \sum_{s=1}^{K_N} \left[
F(\tilde{\beta}_\mathrm{c}(N,s)) - H_{N}(\tilde{\beta}_\mathrm{c}(N,s)) \right
]^2\,.
\end{equation}
In order to get an error estimate we repeat the procedure for $1000$
bootstrap samples for each value of $N$.  In fig.~\ref{fig:TW2} we
consider the theoretical hypothesis of a Gaussian distribution.  We
show in the top part of the figure $\alpha(N)$ versus $N^{-1/3}$, and
our best fit including the first scaling corrections. In the middle
frame we show $\gamma(N)$ versus $N^{-1/3}$, and our best fit. Here
the fitting function only includes the leading term, since this form
already gives a good value for $\chi^2$.  In the bottom frame we show
the collapse of the EDF for different $N$ values and of the
theoretical distribution function, as described in the text. In
fig.~\ref{fig:TW3} we show the same data for the hypothesis of a
TW distribution. 

In short, even if the Gaussian hypothesis is slightly favoured over
the TW one, this analysis does not allow us to decide clearly in one
sense or in the other one. The values of $D$ are always comparable
among the two cases. The estimates of $\gamma(N)$ are indeed more
consistent for the Gaussian case, but the difference of the quality of
the two fits does not allow us to say a clear, final word.

In order to try to sharpen our analysis we have used the Cramer-von
Mises criterion \cite{Anderson,Darling,Darling-par}. We will
not give here technical details (see, however, endnote~\cite{noteCvM}),
but will only discuss the most important features and the results.
When we normalise the distribution to zero average we and unitary
width we introduce a correlations about the values to be tested: also
because of that we find a better fit to our needs the two sample
formulation of the criterion, with a non-parametric approach, where we
have to start by fitting the test statistics (since we cannot use
tabulated values, because we are determining $\alpha_N$ and $\gamma_N$
in Eq.~\ref{eq:wedoitbetter} from our finite-size statistics).  Again,
as in our previous analysis, tests do not allow us to select a Gaussian
or a TW distribution: they are both characterised by very similar
levels of significance.

\begin{figure}
\includegraphics[height=0.7\linewidth,angle=0]{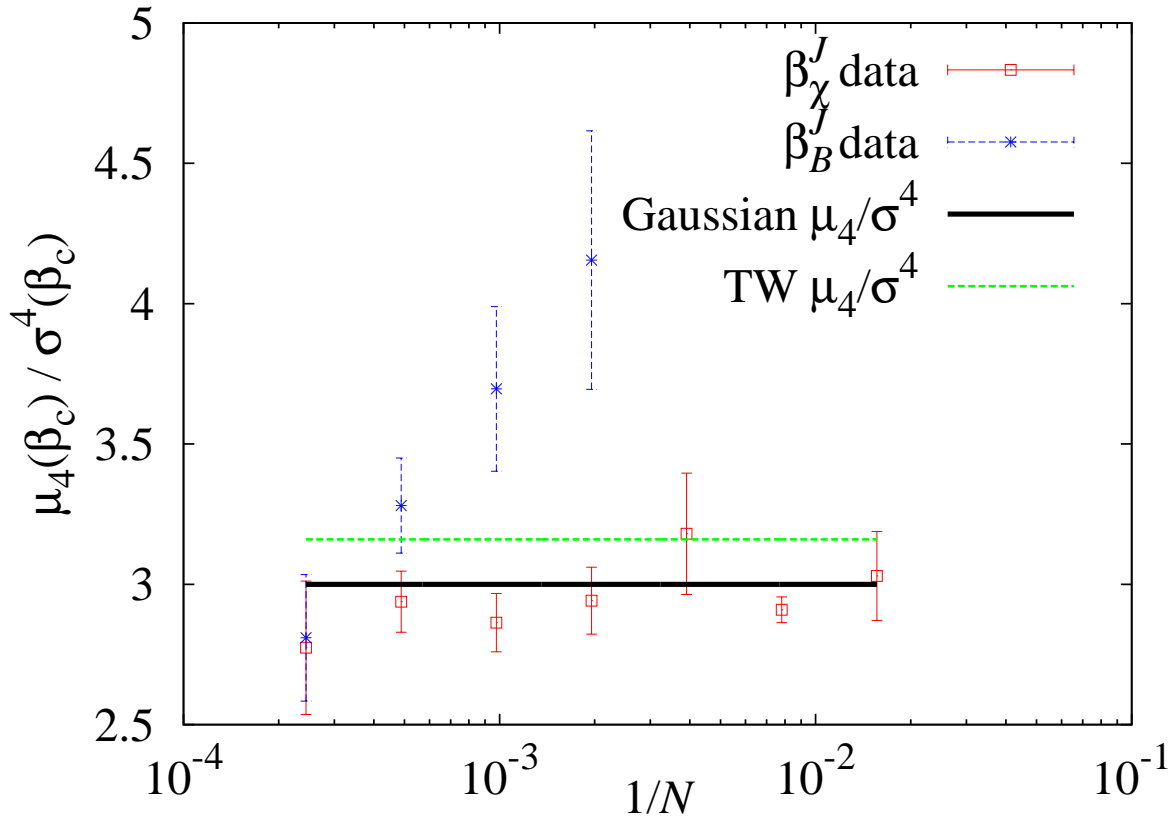}
\caption{Measured fourth moment of the probability distribution, both
  for data from $\chi_{SG}$ and for data from the pseudo Binder
  parameter, versus $1/N$. Thick straight line is for a Gaussian
  distribution (where the value is three), while the thinner straight
  line is for a TW distribution.}
\label{fig:TW4}
\end{figure}

A very simple analysis is maybe the most revealing. As we have
discussed at the start of this section the fourth moment of a
normalised Gaussian is equal to three, while the fourth moment of a
TW distribution is equal to $3.165$. We plot in fig.~\ref{fig:TW4}
the measured fourth moment of the probability distribution, both for
data from $\chi_{SG}$ and for data from the pseudo Binder parameter,
versus $1/N$. The thick straight line is for a Gaussian distribution
(where the value is three), while the thinner straight line is for a
TW distribution. Again, these data do not allow for a precise
statement, but they seem to favour the possibility of a Gaussian
behaviour (the data for $\chi_{SG}$ give maybe the clearer indication).

Our conclusions is that, given the quality of our data and the sizes
of our thermalised configurations, that do not go beyond $N=4096$, a
Gaussian distribution is favoured, but we cannot give a clear,
unambiguous answer.

\section{Conclusions}\label{sect:conclusions}

We have presented a simple method to study the probability
distribution of the pseudo-critical temperature for spin glasses. We
have applied this method to the $3d$ EA Ising spin glass and to the
fully connected SK models. Our results are in excellent agreement with
a median of the distribution that behaves asymptotically like
$T_\mathrm{c}+{\cal O} (L^{-1/\nu })$ (or $1+{\cal O} (N^{-1/3 })$ for
the SK model), and a width of the distribution that behaves like
${\cal O} (L^{-1/\nu })$ (or $1+{\cal O} (N^{-1/3 })$ for the SK
model). The value of $\nu$ we find for the EA model is compatible with
state of the art results. Furthermore, even if our number of samples
is modest as compared with Ref.~\cite{hasenbusch:08b}, our
determination of $\nu$ and $T_\mathrm{c}$ is competitive.  An analysis
of the probability distribution of the pseudo-critical inverse
temperatures for the SK mean field model does not lead to firm
conclusions, but hints to a Gaussian behaviour.

\acknowledgments 

We are indebted to the Janus collaboration that has allowed us to use
equilibrium spin configurations of the $D=3$ Edwards-Anderson
model~\cite{janus:10,janus:10b} obtained by large scale numerical
simulations.  AB thanks C\'ecile Monthus and Thomas Garel for
discussions at an early stage of the work and, specially, Barbara
Coluzzi for a sustained collaboration on the study of the SK model.
We acknowledge partial financial support from MICINN, Spain, (contract
no FIS2009-12648-C03), from UCM-Banco de Santander (GR32/10-A/910383)
and from the DREAM Seed Project of the Italian Institute of Technology
(IIT).  DY was supported by the FPU program (Spain).



\begin{thebibliography}{99.}
\bibitem{Foot1} We are assuming that $\langle S_x\rangle_J\neq 0$ in
  the low temperature phase, i.e. for example that we are working with
  an infinitesimal magnetic field.
\bibitem{noteCvM} Given two independent samples of $N$ and $M$ values
  respectively of a random variable $x$, whose empirical distribution
  functions are $F_N(x)$ and $G_M(x)$, and being $H_{N+M}(x)$ the
  empirical distribution function of the sample obtained by combining
  the two original samples together, the two-samples Cram\'er-von
  Mises statistics is the distance
  $T(N,M)=\left[NM/\left(N+M\right)\right]
  \int_{-\infty}^{\infty}\left[F_N(x)-G_M(x)\right]^2dH_{N+M}(x)$.
\end{thebibliography}
\end{document}